\documentclass[final,3p,times]{elsarticle}

\usepackage{amssymb}
\usepackage{stfloats}
\usepackage{gensymb}
\usepackage{xcolor}
\usepackage{amsmath}

\usepackage{graphicx}
\graphicspath{{images/}} 

\journal{Nuclear Physics B}

\begin{document}

\begin{frontmatter}

\title{Scalable Fabrication of Diamond-on-Silica Heterostructures via High-Selectivity Deep ICP-RIE and Room-Temperature Bonding}

\author{R. Chembra Vasudevan, A. Hammouti, J. Le Pouliquen, T. Batte, P. Pirasteh, Y. Dumeige and P. Huillery} 

\affiliation{organization={Univ Rennes, INSA Rennes, CNRS, Institut FOTON - UMR 6082},
            postcode={F-35000},
            city={Rennes},
            country={France}}

\begin{abstract}
Single-crystal diamond is a leading material platform for high-power electronics and solid-state quantum technologies, yet many device architectures require micrometer-scale membranes with deeply etched features, patterned from commercially available substrates. In this work, we demonstrate a complete through-etch of a 16 $\,\mu\text{m}$ -thick NV-doped single-crystal diamond membrane using a single-layer SiO$_{2}$ hard mask combined with a multi-step oxygen-based ICP-RIE process. With a diamond-to-SiO$_{2}$ selectivity of 15:1, this non-metallic mask strategy can achieve etch depths of few tens of $\,\mu\text{m}$ with well-defined sidewalls, conserved surface roughness and negligible micromasking. Furthermore, we use the oxide layer that remains after etching to serve as the bonding surface in a subsequent integration step. The etched microstructures are transferred onto SiO$_{2}$ substrates and bonded at room temperature using O$_{2}$ plasma surface activation and a sodium silicate interlayer. The resulting siloxane film is optically transparent across the visible spectrum and introduces no detectable parasitic photoluminescence, preserving the optical readout of the embedded NV centers. Together, this deep-etch and room-temperature bonding process provides a scalable and contamination-free route from bulk diamond membranes to diamond-on-silica heterostructures for integrated quantum photonics and sensing applications.
\end{abstract}

\begin{keyword}
Nano and micro-fabrication \sep Diamond etching \sep Diamond-on-silica heterostructures \sep Diamond NV center \sep Quantum sensing
\end{keyword}

\end{frontmatter}

\section{Introduction}
\label{sec1}

Single-crystal diamond has emerged as a promising material platform for integrated photonics and solid-state quantum technologies. Its combination of an ultrawide bandgap \cite{Mingfei}, exceptional thermal conductivity and remarkable hardness makes it well suited for operation in extreme environments \cite{Thomas1993OpticalPO}. Beyond these intrinsic properties, diamond has the ability to host optically active point defects such as the nitrogen-vacancy (NV) center that provides a robust spin-photon interface even at room temperature \cite{Gruber1997, Shandilya2022Diamond}. Together, these properties enable a broad range of applications for quantum sensing, communication and information processing, as well as for advanced photonics, optoelectronics, and high-power devices operating under extreme conditions.\\
Realizing some of these applications requires high-quality single-crystal diamond material and future device architectures will greatly benefit from thin membrane pieces of tens to hundreds $\,\mu\text{m}^2$ area with well-defined geometries, etched from larger commercially available substrates. Examples include compact X-ray and particle-radiation detectors, fiber-based endoscopic magnetometers \cite{Kuwahata2020Magnetometer, hatanoHighprecisionRobustMonitoring2022, vindolet_high-resolution_2025, newman_endoscopic_2025}, high-pressure windows \cite{Piracha}, micro-electromechanical (MEMS) resonators \cite{Sun2023Diamond} or integrated quantum photonic platforms \cite{challier_advanced_2018, Ruf2019Optically}. Achieving these demands fabrication methods that combine high etching depth with strict control over surface morphology. However, diamond's hardness and chemical inertness make it challenging.\\

Current structuring approaches include laser cutting using ultrashort pulsed lasers \cite{Polikarpov2016DiamondXray}, and plasma-based etching processes when precise geometries and tight dimensional control are required. However, fully etching through commercially available membranes, typically 10-20 $\,\mu\text{m}$ thick, remains challenging \cite{Toros2018Precision, zhao_experimental_2025} as it must simultaneously achieve high etch depth, good sidewall quality, and minimal micromasking to preserve optical and mechanical performance.\\
Several methods have been developed, including plasma-based reactive ion etching (RIE), inductively coupled plasma RIE (ICP-RIE), ion and electron beam approaches, and multi-step hybrid processes, each involving trade-offs between etch rate, anisotropy, and surface damage \cite{TOROS2020}. Among these, ICP-RIE has emerged as the dominant technique for deep diamond structuring, as it combines relatively high etch rates with good directional control. ICP-RIE of diamond typically employs oxygen-based, fluorine-based, or chlorine-based plasma chemistry, either pure or mixed with inert gases such as argon to tune the balance between chemical and physical etching \cite{pearton_plasma_2020}. Regardless of the chemistry, achieving etch depths beyond a few micrometers requires hard masks with sufficient selectivity to withstand prolonged plasma etching \cite{TOROS2020}.\\
Hard masks for deep diamond etching generally fall into two categories; metallic and dielectric. Metallic masks such as aluminum, nickel, or chromium are widely used and can provide high selectivity. Under SF$_6$ plasma, selectivities of 11 to Al and 75 to Ni have been reported \cite{Golovanov}. However, extended plasma exposure may cause sputtering and redeposition of metal atoms onto the etched surface and sidewalls, potentially leading to micromasking and increased roughness \cite{TRAN2010778}. To mitigate this effect, oxygen-based plasmas are often combined with secondary gases such as Ar, Cl$_2$, or CF$_4$, which help balance the chemical etch component and reduce micromasking originating from the mask or chamber surfaces. Cycling etch conditions has also been explored as a strategy to limit micromask formation \cite{WANG20246559}. Hicks et al. has for example reported diamond etch depths exceeding 10 $\,\mu\text{m}$  with near-zero micromasking using an optimized cyclic Ar/O$_2$-Ar/Cl$_2$ ICP-RIE process and an Al mask \cite{Hicks2019Diamond}.\\
Dielectric masks, particularly silicon dioxide (silica), offer an attractive alternative to metallic ones as they generally produce cleaner surfaces with reduced redeposition artifacts \cite{TOROS2020}. In the last decade, Toros et al. achieved etch depths up to 150 $\,\mu\text{m}$ for precision micromechanical components using a 7 $\,\mu\text{m}$ thick SiO$_2$ hard mask under high-intensity O$_2$ plasma, with a selectivity of 50:1 and good sidewall quality \cite{Toros2018Precision}. Recently, Corazza et al. demonstrated deep etching exceeding 45 $\,\mu\text{m}$ from bulk diamond using alternating Ar/Cl$_2$ and O$_2$ plasmas with a 10-22 $\,\mu\text{m}$ thick SiO$_2$ mask, producing submicrometer-thick membranes with high surface quality \cite{Corazza}. To achieve such deep etching, SiO$_2$ masks thicker than usually encountered in microfabrication processes have been employed, requiring particular deposition methods such as the use of adhesion layers\cite{Toros2018Precision}. Nevertheless, these results highlight the growing progress of dielectric mask strategy for deep diamond structuring.\\ 

In this work, we employ such a dielectric mask strategy to etch entirely through a commercially sourced 1.5 x 1.5 mm$^2$, 16 $\,\mu\text{m}$ -thick NV-doped single-crystal diamond membrane. The etching process relies on a 2 $\,\mu\text{m}$ single-layer SiO$_2$ hard mask deposited via Plasma-Enhanced Chemical Vapor Deposition (PECVD) and a multi-step oxygen-based ICP-RIE recipe. We succeed to etch entirely through the 16 $\,\mu\text{m}$ membrane with a selectivity of 15:1, sidewalls angle of around 20\degree, conserving the initial diamond surface roughness and with negligible micromasking.\\ 
The use of a dielectric mask compared to metallic ones eliminates the associated sputtering, redeposition, and contamination, which are particularly detrimental to NV-doped diamond intended for quantum applications. Moreover, complete through-etching enables the partitioning of a single commercial membrane into many individual microstructures, reducing the cost per device and providing a scalable route to diamond-based components through heterogeneous or hybrid integration.\\
As a first step in this direction, we successfully transfer single pieces of our etched diamond membrane onto SiO$_2$ substrates to form diamond-on-silica material heterostructures. Transfer is performed by deterministic pick-and-place using vacuum tool and PDMS stamps. Permanent bonding is achieved at room temperature through a sodium silicate interlayer, based on hydroxide catalysed bonding (HCB). This technique was originally developed to adhere optical components for gravity probe B mission. Interestingly, this approach requires no high-temperature annealing and no pressure-assisted bonding equipment. The bonding of silica to silica using this technique has been extensively studied by Elliffe et al. with sodium silicate solution with various dilution factors \cite{elliffeHydroxidecatalysisBondingStable2005}. Bonding have also been realized between different materials supporting silicate-like network on their surfaces such as BK7 to BK7, BK7 to SiC (silicon carbide) and SiC to SiC. With reported bond strength exceeding 2 MPa, the technique was employed for many space mission-based applications to fabricate optical components that withstand harsh launch conditions \cite{prestonHydroxideBondingStrengthMeasurements2008, BondingSiCtoSiC}. Up to our knowledge, this technique has never been used for bonding diamond to other substrates. In this work, we employ it to bond etched diamond pieces on SiO$_2$ substrate, utilizing the remaining layer of SiO$_2$ mask that is deposited on diamond before etching. Remarkably, the interlayer remains optically inert in the visible spectrum. This is important for quantum application as any parasitic fluorescence would degrade the readout of the embedded NV centers.\\
Together, deep dielectric-mask etching and room-temperature silicate bonding provide a complete and contamination-free fabrication strategy to go from bulk NV-doped diamond material to an integrated diamond-on-silica platform for quantum photonics and sensing applications.

\section{Multistep ICP-RIE Etching Process}

The starting material of our work is a 1.5 x 1.5 mm$^2$, 16 $\,\mu\text{m}$ -thick single-crystal diamond membrane with isotopically pure carbon and a NV centers concentration of approximately 1 ppm.\\
Prior to etching, the membrane is cleaned by immersion in a piranha solution (3:1 H$_{2}$SO$_{4}$:H$_{2}$O$_{2}$) at 75 $^\circ$C for 45 minutes, rinsed sequentially in acetone and isopropanol, and dried under nitrogen. A 2 $\,\mu\text{m}$-thick SiO$_2$ hard-mask layer is then deposited onto the cleaned diamond surface by PECVD. The operating temperature for the deposition is 280 $^\circ$C, with gas flow ratios of SiH$_4$:N$_2$O at 40/1000 sccm. The stress value of SiO$_2$ deposited on Si wafer by this recipe is 100 MPa, measured via profilometry. Stress measurements could not be performed for SiO$_2$ deposited on diamond due to the unavailability of diamond substrates larger than few mm$^2$.

\begin{figure*}[h]
  \centering
\includegraphics[width=0.8\textwidth]{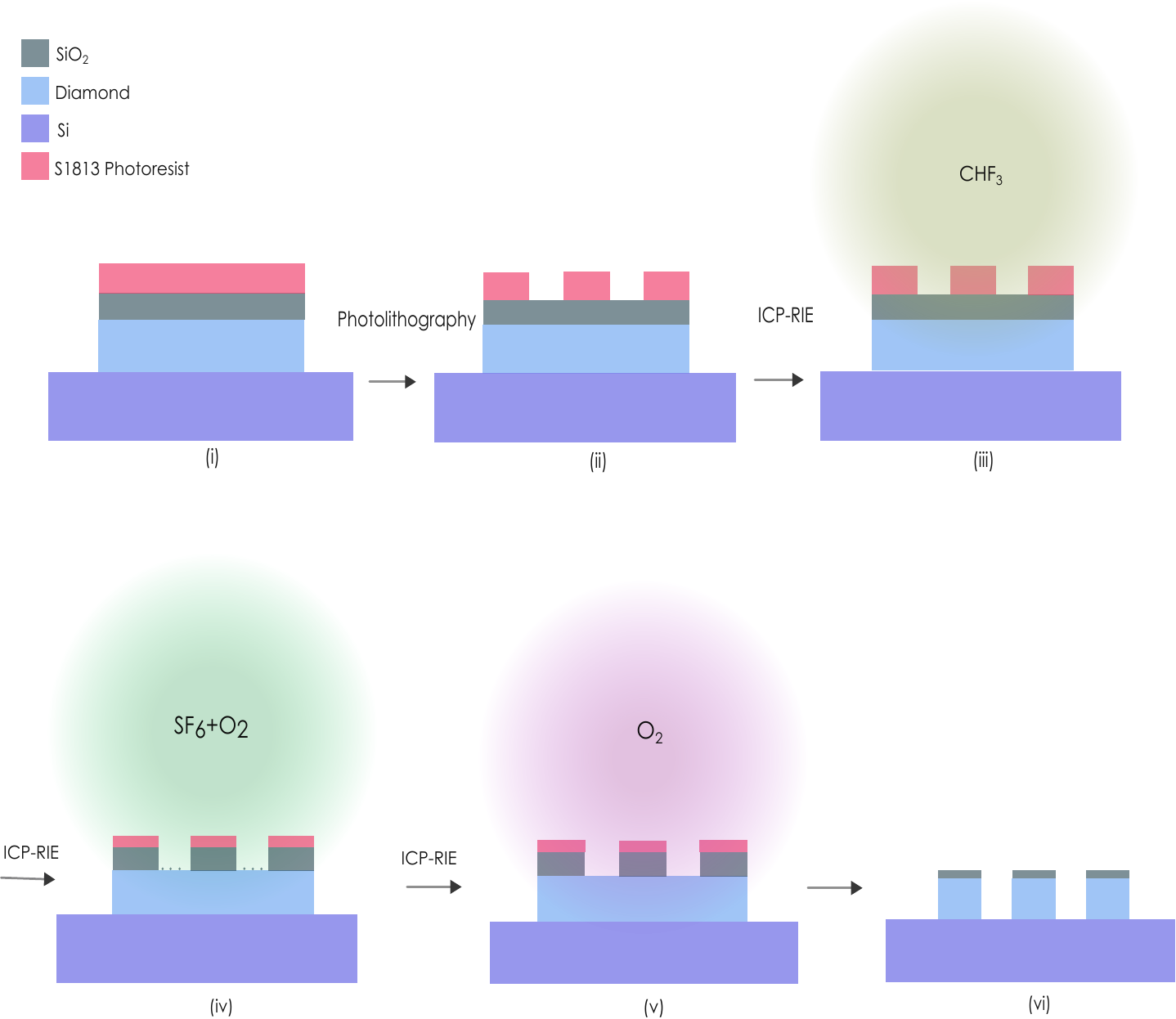}
\caption{\textbf{Schematic overview of the diamond patterning and etching process.} (i) Diamond substrate with deposited SiO$_2$ layer and  spin-coated photoresist attached to a silicon carrier wafer. (ii) Pattern definition in the photoresist using photolithography. (iii) Transfer of the pattern into the SiO$_2$ layer via ICP-RIE using CHF$_3$ plasma. (iv) Brief cleaning step of SF$_6$+O$_2$ plasma. (v) Transfer of the pattern into the diamond substrate via ICP-RIE using O$_2$ plasma. (vi) Final diamond sample on the silicon carrier, with the remaining SiO$_2$ hard mask still present on the surface.}
  \label{fig:process}
\end{figure*}

For handling during lithography and etching, the SiO$_2$ -coated diamond membrane is mounted on a silicon carrier wafer using an acetone soluble adhesive, QuickStick$^{\text{TM}}$ 135. First, the QuickStick is dissolved in acetone and spin-coated onto the silicon wafer to form a uniform layer which is 1 to 2 $\,\mu\text{m}$ thick. The diamond membrane is then placed directly on top of the spin-coated area. Next, the entire assembly is kept on a hotplate at a temperature above 100 $^\circ$C ( flow point is 70 $^\circ$C). At this stage, the adhesive reflows and spreads evenly beneath the diamond membrane. Finally, the region outside the diamond membrane is washed out using a gentle acetone flow for just a few seconds, leaving behind on the wafer only the adhesive layer situated directly under the diamond.\\

As illustrated on Fig. 1, etch pattern is defined by UV photolithography in a S1813 positive photoresist layer spin coated on the sample. The etching process is then performed in three sequential steps. The pattern is first transferred from the S1813 resist into the SiO$_2$ layer using a CHF$_3$ plasma at 5 mTorr, ICP power of 400 W and RF power of 25 W. The SiO$_2$ etch rate under these conditions is 177 nm/min, with a selectivity of 5.2:1 to the S1813 resist. Although this chemistry provides efficient oxide removal, it can leave silica residues at the bottom of the etched features which can act as micromasks during the subsequent diamond etch. A short SF$_6$+ O$_2$ plasma cleaning step is therefore performed to remove these residues and expose a clean diamond surface prior to etching.\\
The diamond itself is finally etched in a pure O$_2$ plasma at 5 mTorr, ICP power of 400 W and RF power of 100 W, yielding an etch rate of 166 nm/min. To prevent overheating over the long process duration required to etch through 16 $\,\mu\text{m}$, the etch is divided into successive 10-minute cycles. Under these conditions, the SiO$_2$ mask eroded at 11.3 nm/min, corresponding to a diamond-to-SiO$_2$ selectivity of 15:1.\\

This multi-step etching process uses gases (CHF$_3$, SF$_6$, O$_2$) of electronic grade (Purity $\geq$ 99.999 \%, total impurities < 10 ppm) and is performed on a Corial 200I ICP-RIE system (Corial Plasma-Therm) with ion energy controlled by a 13.56 MHz RF platen generator. The ICP and RF matching networks are automatically tuned during each run, keeping the reflected power below 1 W and the resulting DC self-bias is monitored in situ. The platen is held at 15 $^\circ$C by a recirculating chiller and helium backside cooling at 20 mTorr to ensure efficient thermal contact to the cathode.\\
We summarized in Table 1 the process parameters used in the three etching steps.

\begin{table}[!h] 
\begin{center}
\begin{tabular}{|l|| c |c |c|} 
 \hline
 Etched material & SiO$_2$ & Residue removal & Diamond \\
 \hline
 Gas & CHF$_3$ & SF$_6$/O$_2$ & O$_2$ \\ 
 \hline
 Gas flow (sccm) & 20 & 25/5 & 20 \\
 \hline
 Pressure (mTorr) & 5 & 20 & 5 \\
 \hline
 RF power (W) & 25 & 100 & 100 \\
 \hline
 ICP power (W) & 400 & 0 & 400 \\
 \hline
 RF bias (V) & 44 & 280 & 185 \\
 \hline
\end{tabular}
\caption{ICP-RIE parameters for the three sequential steps of the etching process.}
\end{center} 
\end{table}

This etching recipe has been tested on few diamond samples prior to be applied to the NV-doped membrane. While the process is qualitatively reliable and reproducible, we observed some variation in the diamond etch rate and the diamond-to-SiO$_2$ selectivity with values ranging respectively from 106 to 220 nm/min and from 10:1 to 19.5:1. We attribute this variation to different effective temperature of diamond samples during etching caused by uneven thermal contact to the reactor's platen.\\ 

\begin{figure*}[h]
  \centering
   \includegraphics[width=0.75\textwidth]{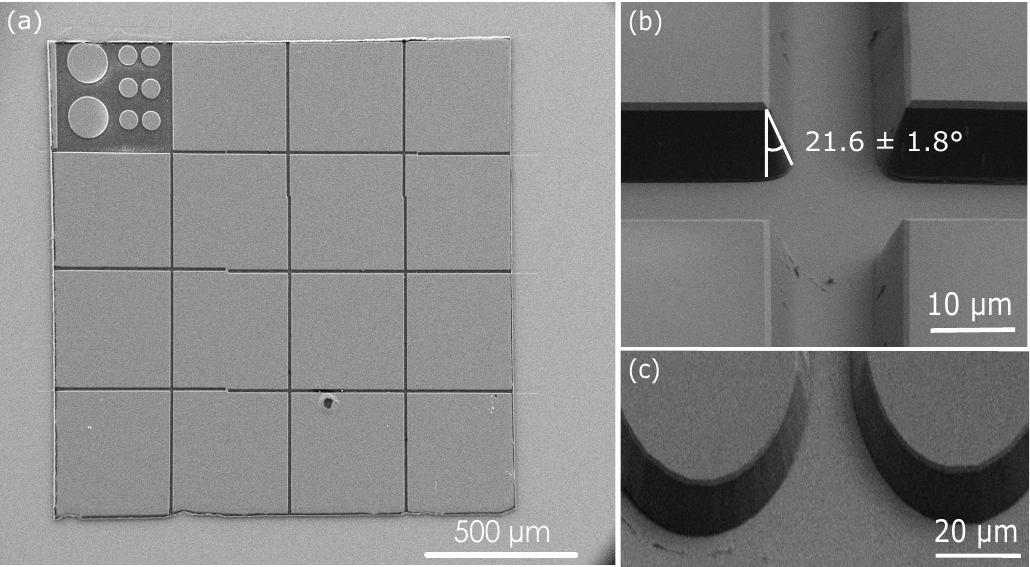}
 \caption{\textbf{Characterization of the etched diamond membrane.} (a) SEM image of the diamond membrane after the etching process. (b,c) SEM images of some regions of interest of the etched membrane, viewed at a 35$^\circ$ tilt angle. Sidewall angle is overlaid on panel (b).}
  \label{fig:topo}
\end{figure*}

As visible in the Scanning Electron Microscopy (SEM) image shown on Fig. 2a, the complete through-etching of the diamond membrane allows us to fabricate diamond pieces with well defined geometries, ready for further integration.\\  
As shown on Fig. 2b and c, our process leads to a high aspect-ratio etching with negligible micromasking. Sidewall angle has been measured from analyzing SEM images of multiple sidewalls and yields a value of 21.6 $\pm$ 1.8 $^\circ$. The surface roughness of the etched diamond sidewalls has been measured with Atomic Force Microscopy (AFM) yielding a value for the RMS roughness of 3.1 $\pm$ 0.2 nm. This appears to be slightly smaller than the roughness of the bare diamond surface measured to be 3.94 $\pm$ 0.04 nm. This shows that our process do not degrade the surface roughness, at least when the latter is in the few nm range. Details about AFM measurements are given in the supplementary material.

\begin{figure*}[h]
\centering
\includegraphics[width=0.8\textwidth]{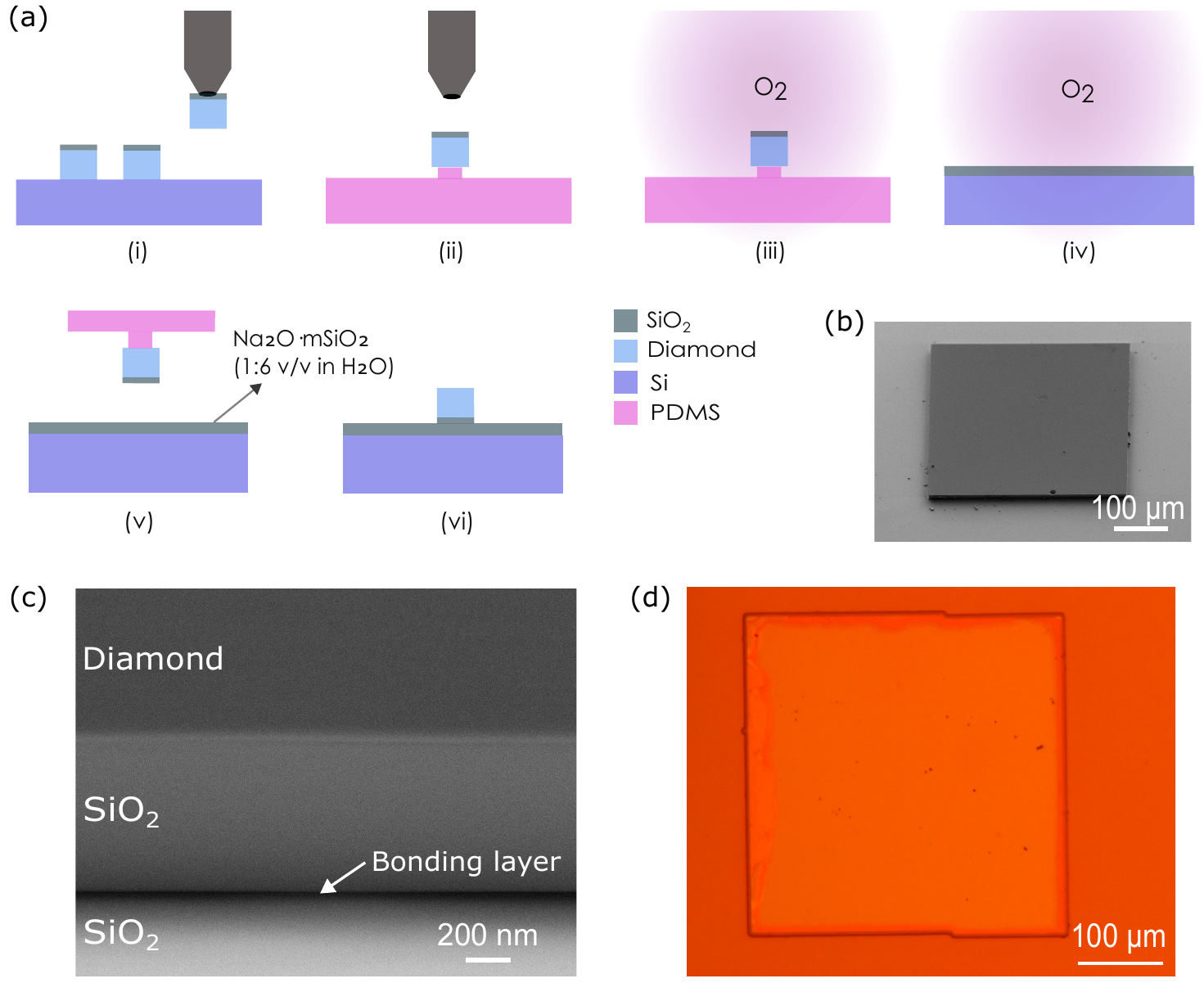}
\caption{\textbf{Room-Temperature Sodium Silicate Bonding.} (a) Schematic illustration of the room-temperature sodium silicate bonding sequence. (i-ii) Transfer of a membrane piece onto a PDMS stamp using a vacuum tool. (iii) Activation of the silica surface of the diamond membrane by exposure to an O$_2$ plasma. (iv) Activation of the silica surface of the host substrate by exposure to an O$_2$ plasma. (v) The diamond membrane is brought in contact with the activated substrate after spin coating of the latter with the sodium silicate solution. (vi) The assembly is stored in a desiccator for four days at room temperature.(b) SEM image of a bonded diamond membrane. (c) Cross-sectional SEM image of the bonded interface. (d) Optical microscope image of a bonded diamond membrane.}
    \label{fig:process_and_characterization}
\end{figure*}

\section{Diamond-on-silica Heterostructures via Room-Temperature Sodium Silicate Bonding}
\label{sec:fabrication}

After the etching process, a 1.2 $\,\mu\text{m}$ thick layer of SiO$_2$ remains on the diamond surface. The roughness of the SiO$_2$ layer after etching is measured to be 0.67 $\pm$ 0.15 nm which appears to be smaller than the 2.4 $\pm$ 0.1 nm roughness of the SiO$_2$ layer deposited on the diamond before etching. While this layer could be readily removed with appropriate RIE process, we use it to bond etched diamond membranes onto silica coated substrates, e.g. to fabricate diamond-on-silica heterostructures.\\
The process we developed is illustrated on Fig 3a. For this, the SiO$_2$ mask layer remaining on the diamond membrane after the deep etch, serves a dual function: it provides a high-quality optical interface with the diamond membrane and a chemically compatible surface for bonding on a host silica substrate. 

\subsection{Surface Preparation and Plasma Activation}
\label{sec:surface_prep}

To facilitate the manipulation of small and fragile membrane, the diamond with its silica cap is mounted onto a polydimethylsiloxane (PDMS) stamp. We fabricate PDMS stamps from SYLGARD$^{\text{TM}}$ 184 silicone elastomer using silicon wafer molds patterned via pholithography and wet etching using 5 \% tetramethylammonium hydroxide (TMAH) solution. The PDMS is cured at 80 $^\circ$C for 1 hour yielding a stiffness of 2.2 MPa. More details on the PDMS stamps preparation are given in the supplementary material.\\
To detach the etched diamond pieces temporarily mounted on the silicon wafer with QuickStick$^{\text{TM}}$ 135, we place the sample on a hot plate and heat it at 90 $^\circ$C to soften the adhesive. We then use a micromanipulator to slide the desired diamond piece on the silicon wafer over few hundreds of $\,\mu\text{m}$ which effectively removes the adhesive from the diamond. A vacuum tool is finally employed to precisely pick up and position the membrane onto the PDMS stamp without inducing mechanical damage.\\

We activate the surfaces to be bonded by exposing them to an oxygen plasma immediately prior to the bonding to promote hydrophilicity and generate surface silanol (Si-OH) groups, which are critical for the condensation chemistry that follows \cite{matinfarReviewSodiumSilicate2023}. The target substrate, a silicon wafer bearing a 300 nm PECVD SiO\(_2\) layer is activated using an O\(_2\) plasma at an RF power of 50 W, a chamber pressure of 100 mTorr, and an O\(_2\) flow rate of 100 sccm for a duration of 10 minutes. The silica surface on the diamond membrane, while on the PDMS stamp, is activated under identical plasma conditions but for a reduced duration of 1 minute. This shorter exposure avoids the risk of plasma-induced damage to the PDMS stamp or the thin diamond membrane while still ensuring adequate surface activation.

\subsection{Bonding Procedure and Interfacial Chemistry}
\label{sec:bonding}

Following plasma activation, an aqueous sodium silicate solution serves as the intermediate bonding medium. The solution is prepared by diluting sodium silicate, Na\(_2\)O\(\cdot\)mSiO\(_2\) where the modulus \(m\) is typically between 3 and 4, in deionized water at a 1:6 volumetric ratio. Preston et al. have shown that ratios from 1:4 to 1:6 can be used without significant differences in the bond shear breaking strength \cite{prestonHydroxideBondingStrengthMeasurements2008}. This solution is subsequently spin-coated onto the activated SiO\(_2\) substrate, yielding a uniform thin film.\\
The PDMS stamp, carrying the activated diamond membrane, is then inverted and brought into gentle contact with the substrate coated with the sodium silicate film. After a brief time of the order of one minute allowing for interfacial wetting and initial bond formation, the PDMS stamp is carefully peeled away while the diamond membrane remains adhered to the substrate.\\
To achieve full chemical bonding, the sample is finally stored in a desiccator at room temperature for a period of four days. This extended duration allows for the gradual dehydration and polycondensation of the silicate layer.\\

As described by Elliffe et al. \cite{elliffeHydroxidecatalysisBondingStable2005}, the chemistry involved in the sodium silicate bonding takes place in three steps, 1) Hydration and etching, 2) polymerization and 3) dehydration.\\
First, the $\text{OH}^-$ ions in the solutions act as a catalyst, etching the silica surface to liberate silicate ions:
\begin{center}
    \begin{equation}
        \text{SiO}_2 + \text{OH}^- + 2\text{H}_2\text{O} \longrightarrow \text{Si(OH)}_5^-.
        \label{eq:HCB_reaction1}
    \end{equation}
\end{center}
Next, as the concentration of silicate ions increases in the bonding solution, the number of active $\text{OH}^-$ ions is reduced and the pH of the solution is decreased. When the pH of the bonding solution falls below 11, the silicate ion dissociates to form $\text{Si(OH)}_4$:
\begin{center}
    \begin{equation}
        \text{Si(OH)}_5^- \longrightarrow \text{Si(OH)}_4 + \text{OH}^-.
        \label{eq:HCB_reaction2}
    \end{equation}
\end{center}
Finally, $\text{Si(OH)}_4$ molecules combine and polymerize to form siloxane chains and water:
\begin{center}
    \begin{equation}
        2\text{Si(OH)}_4 \longrightarrow (\text{HO})_3\text{SiOSi(OH)}_3 + \text{H}_2\text{O}.
        \label{eq:HCB_reaction3}
    \end{equation}
\end{center}
As the water evaporates, a three dimensional network of tangled siloxane chains joins the silica surfaces resulting in a robust, mechanically stable bond, with breaking stress measured to be in the MPa range\cite{elliffeHydroxidecatalysisBondingStable2005}.\\

In this study, we tried and succeed in all attempts to bond 4 individual membranes with this process. A SEM image of one of the final heterostructure is shown on Fig. 3b. A cross-sectional SEM image of the bonded interface, shown on Fig. 3c, reveals a uniform bonding layer, with a thickness of few dozen of nm.\\
We checked that the bonds are stable enough to withstand different operations relevant for further processing of the heterostructures. For instance, all bonded membranes have been cleaned in piranha solution at 75 $^\circ$C for 45 minutes which is our usual diamond cleaning procedure, some were immersed in hydrofluoric acid (HF) for up to 20 s and others exposed to SF$_6$ plasma for up to 30 min. More details are given in the supplementary material.\\
As shown on Fig. 3d, optical microscope characterization of a bonded membrane reveals a uniform bond between the two silica surfaces over the entire diamond area, apart from the membrane border where silica material was removed after immersion in HF solution. Obtaining such a full contact is critical for minimizing scattering losses in future photonic devices, as it effectively creates a monolithic silica cladding. As shown in the supplementary material, 2 out of the 4 bonded membranes feature void regions localized in one of their corner. We believe this could be avoided by using adequate micromanipulation instead of manual handling when bringing the diamond membrane in contact with the sodium silicate coated wafer. 

\section{Optical characterization of the embedded NV centers}
\label{sec:characterization}

We performed photoluminescence (PL) spectroscopy to asses the impact of the bonding process on the optical readout of the diamond NV centers embedded in the fabricated heterostructures. As illustrated on Fig. 4a, this is done with a confocal microscope where around 30 mW of green (532 nm) laser light is focused on the sample. With a spot size of the order of 50 $\,\mu\text{m}$ and a Rayleigh range exceeding 1 mm, PL is excited and collected from the entire depth of the heterostructures. PL spectra are acquired using a Horiba IHR 320 spectrometer equipped with a sincerity CCD camera.\\

As shown on Fig. 4b, the normalized PL spectrum across the visible range (550 nm to 900 nm) of a processed and bonded membrane is found to be essentially identical to the one of the bare diamond sample. Both spectra were taken with an exposure time of 1s and exhibit similar signal-to-noise ratio. Those PL spectra can be deconvoluted into two spectral responses belonging to (NV\(^0\)) and (NV\(^-\)) centers in diamond. As indicated by the colored area on Fig. 4b, the characteristic zero-phonon lines of both charge states and their respective phonon sidebands are clearly resolved and respectively fitted with Lorentzian and Gaussian functions. On top of that, PL spectra exhibit weak oscillations arising from interference effects in the thin material layers. Similar PL spectra acquired from different locations of another bonded membrane are shown in the supplementary material.\\ 
Importantly, those PL measurements show that no additional fluorescence features attributable to the sodium silicate bonding layer, process contaminants, or optically active interfacial defects are observed within the NV centers emission range. The NV centers charge state is also not affected by the process in a detectable way. This confirms that the etching process, the room-temperature bonding chemistry and the prolonged dehydration step do not degrade the optical emission nor the spectral integrity of the NV centers as probed in the fabricated heterostructures.\\

\begin{figure*}[hthb]
\centering
\includegraphics[width=1\textwidth]{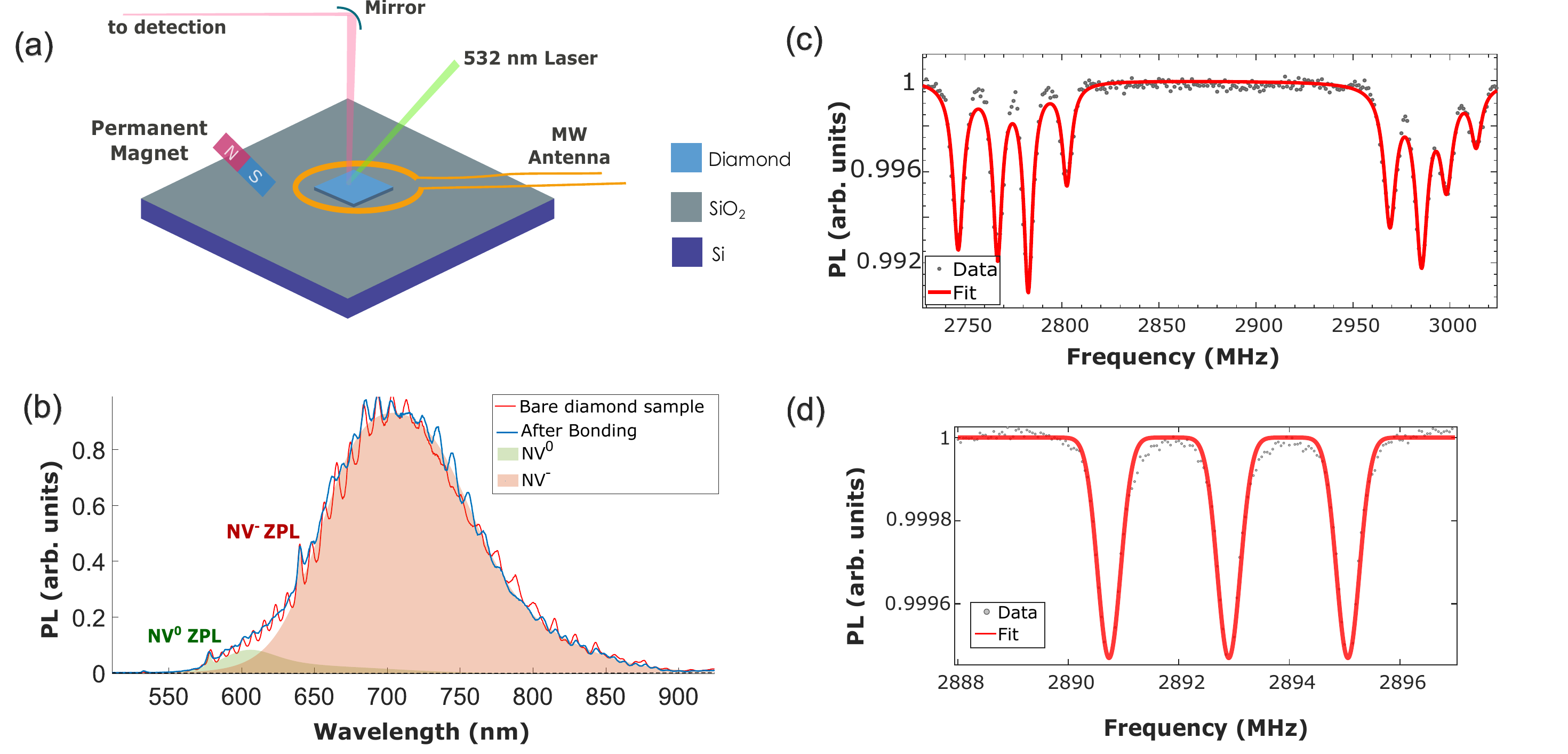}
\caption{\textbf{Characterization of NV centers in the fabricated heterostructures.} 
    \textbf{a)} Scheme of the experimental setup used for the characterization of the heterostructures. 
    \textbf{b)} Photoluminescence (PL) spectra acquired from the bare diamond sample (red solid line) and from a fabricated heterostructure (blue solid line). The heterostructure spectrum is fitted with the sum of 4 functions, a Lorentzian for the zero-phonon line (ZPL) and a Gaussian for the phonon sideband, this for both (NV\(^0\)) and (NV\(^-\)) charge states.
    \textbf{c)} Optically detected magnetic resonance (ODMR) spectrum showing the Zeeman splitting of the spin sublevels. The fit is a guide to the eye. 
    \textbf{d)} ODMR spectrum showing the NV's hyperfine transitions, with Gaussian fits (FWHM: 0.49 $\pm$ 0.01 MHz).}
\label{fig:PL}
\end{figure*}

We further investigate the quantum properties of the embedded NV centers by performing Optically Detected Magnetic Resonance (ODMR) measurements. For this, a permanent magnet and a home-made wire-loop microwave antenna fed by an amplified microwave generator are added to the PL setup (see Fig. 4a). The PL signal is spectrally filtered below 600 nm and above 800 nm using optical filters and detected with a photodiode.\\ 
Fig. 4c shows an ODMR spectrum taken at a B-field of around 7 mT where we observe 8 spin resonances with a contrast close to 1 \%. This is characteristic of a NV-ensemble where NV centers are distributed along the 4 possible crystalline orientations of the diamond lattice. For this measurement, the PL signal was integrated over 3 ms for each microwave frequency within a scan and averaged over 100 scans.\\
The ODMR spectrum shown on Fig. 4d, taken with very small microwave power and a magnetic field of around 2 mT, allows to resolve the NV centers hyperfine spectrum and can be used to evaluate the NV centers coherence time within the heterostructure. For this measurement, the PL signal was integrated over 3 ms for each microwave frequency within a scan and averaged over 250 scans to keep a descent signal-to-noise ratio despite the very small contrast arising from the weak microwave excitation. The lineshape parameters were extracted from a Gaussian fit of the CW-ODMR spectrum, with a full width at half maximum (FWHM) of 0.49 $\pm$ 0.01 MHz. In the limit of very weak microwave excitation, the inhomogeneous dephasing time $T_2^*$ is related to the Gaussian resonance linewidth by the following relationship \cite{PhysRevB.84.195204}:
\begin{equation}
    T_2^* = \frac{2\sqrt{\ln(2)}}{\pi \cdot \text{FWHM}},
\end{equation}
which yields a dephasing time of $T_2^* = 1.07 \pm 0.03 \,\mu\text{s}$. This matches the $T_2^* = 1\,\mu\text{s}$ rated by the supplier for the NV-doped diamond sample used in this work, suggesting that our etching and bonding processes did not significantly affect the NV centers coherence time. 

\section{Conclusion and perspectives}
In conclusion, we have demonstrated an effective and scalable method for fabricating deep-etched single-crystal diamond membranes utilizing a thin 2 $\,\mu\text{m}$-thick silica hard mask. By achieving etch depths exceeding 16 $\,\mu\text{m}$  while maintaining good sidewall quality and eliminating the micromasking issues associated with metallic masks, this technique provides a clean approach for producing high-aspect-ratio diamond microstructures. The subsequent bonding of these membranes onto SiO$_{2}$ via a room-temperature sodium silicate bonding process creates stable diamond-on-silica heterostructures. The absence of detectable parasitic fluorescence in the fabricated heterostructures and the measurement of NV centers coherence time $T_2^*$ exceeding 1 $\,\mu\text{s}$ indicate that our low-cost, plasma-activated bonding technique is well suited for sensitive quantum applications.\\
As future perspectives, the "deep etch and transfer" method we developed grants a marked economic advantage by allowing the scalable distribution of high-quality diamond membranes onto diverse substrates, reducing the material cost per device. Diamond-on-silica heterostructures could be immediately applicable for NV-based quantum sensing, especially to build cost-effective endoscopes with NV-doped diamonds attached to the tip of silica optical fibers or for radiation detectors. Furthermore, the realization of photonic devices such as compact photonic circuits and high quality-factor resonators necessitates the integration of high-refractive-index materials with a low-index cladding to provide strong optical confinement. Diamond, with a refractive index of approximately 2.4 in the visible range, and silica (SiO\(_2\)), with an index of roughly 1.45, form a suitable material pair for such applications. Starting from the heterostructures presented in this work, further thinning of the diamond layer down to sub-micron scale and subsequent photonic structuration would provide an adaptable foundation for integrated diamond photonic circuits and advanced sensing platforms \cite{faraon_resonant_2011}. This forms an alternative solution to monolithic diamond devices fabricated using either laser processing \cite{Giakoumaki} or quasi-isotropic etching \cite{mitchell_realizing_2019} where air is used as cladding, integration of ultra-thin diamond membranes produced with ion-based slicing methods \cite{ding_high-q_2024, guo_direct-bonded_2024} or pick-flip-and-place transfer printing of angle-etched diamond structures \cite{katsumi_hybrid_2023}.\\
In future work, the realization of photonic devices employing laser excitation from the silica side of bonded heterostructures will allow to further check the quality of the bonded interface, namely regarding light scattering that could potentially affect NV centers excitation and PL collection in this configuration. Moreover, probing NV centers close to the bonding interface, for example in sub-micron scale photonic circuits will allow to check on the potential presence of paramagnetic defects at the material interfaces that could induce magnetic noise on NV centers and broaden their lineshape.

\section*{CRediT authorship contribution statement}

\textbf{R. Chembra Vasudevan:} Methodology, Investigation, Data Curation, Writing - Original Draft, Writing - Review \& Editing, Visualization. \textbf{A. Hammouti:} Methodology, Investigation, Data Curation, Writing - Original Draft, Writing - Review \& Editing. \textbf{J. Le Pouliquen:} Methodology, Investigation, Writing - Review \& Editing. \textbf{T. Batte:} Methodology, Investigation, Writing - Review \& Editing. \textbf{P. Pirasteh:} Methodology, Writing - Review \& Editing, Supervision. \textbf{Y. Dumeige:} Conceptualization, Writing - Review \& Editing, Supervision, Funding acquisition. \textbf{P. Huillery:} Conceptualization, Methodology, Writing - Original Draft, Writing - Review \& Editing, Supervision, Funding acquisition.

\section*{Funding sources}

This work has received funding from the French National Research Agency (ANR) with the SINFONIA project (grant number ANR-23-QUAC-0006-01) and the ESR/EquipEx+ program with the e-Diamant project (grant number ANR-21-ESRE-0031) and from Rennes Metropole.

\section*{Aknowledgements}

The authors acknowledge nanoRennes and CCLO for technological support, platforms affiliated to RENATECH+ (the French national facilities network for micro-nanotechnology). The authors would like to thank V. Mille and X. Checoury for useful discussions.

\section*{Declaration of competing interest}

The authors declare that they have no known competing financial interests or personal relationships that could have appeared to influence the work reported in this paper.

\section*{Data availability}

The data that support the findings of this study are available from
the corresponding author upon reasonable request.

\bibliographystyle{elsarticle-num}
\bibliography{mybib.bib}

\begin{thebibliography}{10}
\expandafter\ifx\csname url\endcsname\relax
  \def\url#1{\texttt{#1}}\fi
\expandafter\ifx\csname urlprefix\endcsname\relax\def\urlprefix{URL }\fi
\expandafter\ifx\csname href\endcsname\relax
  \def\href#1#2{#2} \def\path#1{#1}\fi

\bibitem{Mingfei}
M.~Xu, D.~Wang, K.~Fu, D.~H. Mudiyanselage, H.~Fu, Y.~Zhao, A review of
  ultrawide bandgap materials: properties, synthesis and devices, Oxford Open
  Materials Science 2~(1) (2022) itac004.
\newblock \href {https://doi.org/10.1093/oxfmat/itac004}
  {\path{doi:10.1093/oxfmat/itac004}}.

\bibitem{Thomas1993OpticalPO}
M.~E. Thomas, W.~J. Tropf, Optical properties of diamond, in: Optics \&
  Photonics, 1993.

\bibitem{Gruber1997}
A.~Gruber, A.~Drabenstedt, C.~Tietz, L.~Fleury, J.~Wrachtrup, C.~von
  Borczyskowski, Scanning confocal optical microscopy and magnetic resonance on
  single defect centers, Science 276~(5321) (1997) 2012--2014.
\newblock \href {https://doi.org/10.1126/science.276.5321.2012}
  {\path{doi:10.1126/science.276.5321.2012}}.

\bibitem{Shandilya2022Diamond}
P.~K. Shandilya, S.~Flagan, N.~Carvalho, E.~Zohari, V.~K. Kavatamane, J.~Losby,
  P.~Barclay, Diamond integrated quantum nanophotonics: Spins, photons and
  phonons, Journal of Lightwave Technology 40 (2022) 7538--7571.
\newblock \href {https://doi.org/10.1109/jlt.2022.3210466}
  {\path{doi:10.1109/jlt.2022.3210466}}.

\bibitem{Kuwahata2020Magnetometer}
A.~Kuwahata, T.~Kitaizumi, K.~Saichi, T.~Sato, R.~Igarashi, T.~Ohshima,
  Y.~Masuyama, T.~Iwasaki, M.~Hatano, F.~Jelezko, M.~Kusakabe, T.~Yatsui,
  M.~Sekino, Magnetometer with nitrogen-vacancy center in a bulk diamond for
  detecting magnetic nanoparticles in biomedical applications, Scientific
  Reports 10 (2020) 2483.
\newblock \href {https://doi.org/10.1038/s41598-020-59064-6}
  {\path{doi:10.1038/s41598-020-59064-6}}.

\bibitem{hatanoHighprecisionRobustMonitoring2022}
Y.~Hatano, J.~Shin, J.~Tanigawa, Y.~Shigenobu, A.~Nakazono, T.~Sekiguchi,
  S.~Onoda, T.~Ohshima, K.~Arai, T.~Iwasaki, M.~Hatano, High-precision robust
  monitoring of charge/discharge current over a wide dynamic range for electric
  vehicle batteries using diamond quantum sensors, Scientific Reports 12~(1)
  (2022) 13991.
\newblock \href {https://doi.org/10.1038/s41598-022-18106-x}
  {\path{doi:10.1038/s41598-022-18106-x}}.

\bibitem{vindolet_high-resolution_2025}
B.~Vindolet, B.~Ducharne, H.~N. Nguyen, X.~Mougenot, C.~Gallais, T.~Hingant,
  High-resolution non-destructive detection of grinding burns with {NV} diamond
  quantum magnetometer, NDT \& E International 155 (2025) 103439.
\newblock \href {https://doi.org/10.1016/j.ndteint.2025.103439}
  {\path{doi:10.1016/j.ndteint.2025.103439}}.

\bibitem{newman_endoscopic_2025}
A.~Newman, S.~Graham, C.~Stephen, A.~Edmonds, M.~Markham, G.~Morley, Endoscopic
  diamond magnetometer for cancer surgery, Physical Review Applied 24~(2)
  (2025) 024029.
\newblock \href {https://doi.org/10.1103/znt3-988w}
  {\path{doi:10.1103/znt3-988w}}.

\bibitem{Piracha}
A.~H. Piracha, K.~Ganesan, D.~W.~M. Lau, A.~Stacey, L.~P. McGuinness,
  S.~Tomljenovic-Hanic, S.~Prawer, Scalable fabrication of high-quality{,}
  ultra-thin single crystal diamond membrane windows, Nanoscale 8 (2016)
  6860--6865.
\newblock \href {https://doi.org/10.1039/C5NR08348F}
  {\path{doi:10.1039/C5NR08348F}}.

\bibitem{Sun2023Diamond}
H.~Sun, Z.~Zhang, Y.~Liu, G.~Chen, T.~Li, M.~Liao, Diamond mems: From classical
  to quantum, Advanced Quantum Technologies 6 (2023).
\newblock \href {https://doi.org/10.1002/qute.202300189}
  {\path{doi:10.1002/qute.202300189}}.

\bibitem{challier_advanced_2018}
M.~Challier, S.~Sonusen, A.~Barfuss, D.~Rohner, D.~Riedel, J.~Koelbl,
  M.~Ganzhorn, P.~Appel, P.~Maletinsky, E.~Neu, Advanced {Fabrication} of
  {Single}-{Crystal} {Diamond} {Membranes} for {Quantum} {Technologies},
  Micromachines 9~(4) (2018) 148.
\newblock \href {https://doi.org/10.3390/mi9040148}
  {\path{doi:10.3390/mi9040148}}.

\bibitem{Ruf2019Optically}
M.~Ruf, M.~IJspeert, S.~V. van Dam, N.~de~Jong, H.~van~den Berg, G.~Evers,
  R.~Hanson, Optically coherent nitrogen-vacancy centers in micrometer-thin
  etched diamond membranes, Nano Letters 19 (2019) 3987 -- 3992.
\newblock \href {https://doi.org/10.1021/acs.nanolett.9b01316}
  {\path{doi:10.1021/acs.nanolett.9b01316}}.

\bibitem{Polikarpov2016DiamondXray}
M.~Polikarpov, V.~Polikarpov, I.~Snigireva, A.~Snigirev, Diamond x-ray
  refractive lenses with high acceptance, Physics Procedia 84 (2016) 213--220.
\newblock \href {https://doi.org/10.1016/j.phpro.2016.11.040}
  {\path{doi:10.1016/j.phpro.2016.11.040}}.

\bibitem{Toros2018Precision}
A.~Toros, M.~Kiss, T.~Graziosi, H.~Sattari, P.~Gallo, N.~Quack, Precision
  micro-mechanical components in single crystal diamond by deep reactive ion
  etching, Microsystems \& Nanoengineering 4 (2018).
\newblock \href {https://doi.org/10.1038/s41378-018-0014-5}
  {\path{doi:10.1038/s41378-018-0014-5}}.

\bibitem{zhao_experimental_2025}
L.~Zhao, X.~Wang, N.~Jiang, K.~Nishimura, J.~Yi, S.~Fang, Experimental study of
  inductively coupled plasma etching of patterned single crystal diamonds,
  Scientific Reports 15~(1) (2025) 21062, publisher: Nature Publishing Group.
\newblock \href {https://doi.org/10.1038/s41598-025-08066-3}
  {\path{doi:10.1038/s41598-025-08066-3}}.

\bibitem{TOROS2020}
A.~Toros, M.~Kiss, T.~Graziosi, S.~Mi, R.~Berrazouane, M.~Naamoun,
  J.~{Vukajlovic Plestina}, P.~Gallo, N.~Quack, Reactive ion etching of single
  crystal diamond by inductively coupled plasma: State of the art and catalog
  of recipes, Diamond and Related Materials 108 (2020) 107839.
\newblock \href {https://doi.org/https://doi.org/10.1016/j.diamond.2020.107839}
  {\path{doi:https://doi.org/10.1016/j.diamond.2020.107839}}.

\bibitem{pearton_plasma_2020}
S.~J. Pearton, E.~A. Douglas, R.~J. Shul, F.~Ren, Plasma etching of wide
  bandgap and ultrawide bandgap semiconductors, Journal of Vacuum Science \&
  Technology A 38~(2) (2020) 020802.
\newblock \href {https://doi.org/10.1116/1.5131343}
  {\path{doi:10.1116/1.5131343}}.

\bibitem{Golovanov}
A.~V. Golovanov, V.~S. Bormashov, N.~V. Luparev, S.~A. Tarelkin, S.~Y.
  Troschiev, S.~G. Buga, V.~D. Blank, Diamond microstructuring by deep
  anisotropic reactive ion etching, physica status solidi (a) 215~(22) (2018)
  1800273.
\newblock \href {https://doi.org/https://doi.org/10.1002/pssa.201800273}
  {\path{doi:https://doi.org/10.1002/pssa.201800273}}.

\bibitem{TRAN2010778}
D.~Tran, C.~Fansler, T.~Grotjohn, D.~Reinhard, J.~Asmussen, Investigation of
  mask selectivities and diamond etching using microwave plasma-assisted
  etching, Diamond and Related Materials 19~(7) (2010) 778--782, proceedings of
  Diamond 2009, The 20th European Conference on Diamond, Diamond-Like
  Materials, Carbon Nanotubes and Nitrides, Part 2.
\newblock \href {https://doi.org/https://doi.org/10.1016/j.diamond.2010.02.001}
  {\path{doi:https://doi.org/10.1016/j.diamond.2010.02.001}}.

\bibitem{WANG20246559}
X.~Wang, S.~Fang, B.~Wang, M.~Qiu, K.~Nishimura, N.~Jiang, J.~Yi, Diamond
  etching with near-zero micromasking, Journal of Materials Research and
  Technology 33 (2024) 6559--6564.
\newblock \href {https://doi.org/https://doi.org/10.1016/j.jmrt.2024.11.011}
  {\path{doi:https://doi.org/10.1016/j.jmrt.2024.11.011}}.

\bibitem{Hicks2019Diamond}
M.-L. Hicks, A.~Pakpour-Tabrizi, R.~Jackman, Diamond etching beyond 10??m with
  near-zero micromasking, Scientific Reports 9 (2019).
\newblock \href {https://doi.org/10.1038/s41598-019-51970-8}
  {\path{doi:10.1038/s41598-019-51970-8}}.

\bibitem{Corazza}
A.~Corazza, S.~Ruffieux, Y.~Zhu, C.~A. Jaramillo~Concha, Y.~Fontana,
  C.~Galland, R.~J. Warburton, P.~Maletinsky, Homogeneous free-standing
  nanostructures from bulk diamond over millimeter scales for quantum
  technologies, Nano Letters 25~(40) (2025) 14526--14533, pMID: 41005752.
\newblock \href {https://doi.org/10.1021/acs.nanolett.5c03083}
  {\path{doi:10.1021/acs.nanolett.5c03083}}.

\bibitem{elliffeHydroxidecatalysisBondingStable2005}
E.~J. Elliffe, J.~Bogenstahl, A.~Deshpande, J.~Hough, C.~Killow, S.~Reid,
  D.~Robertson, S.~Rowan, H.~Ward, G.~Cagnoli, Hydroxide-catalysis bonding for
  stable optical systems for space, Classical and Quantum Gravity 22~(10)
  (2005) S257.
\newblock \href {https://doi.org/10.1088/0264-9381/22/10/018}
  {\path{doi:10.1088/0264-9381/22/10/018}}.

\bibitem{prestonHydroxideBondingStrengthMeasurements2008}
A.~Preston, B.~Balaban, G.~Mueller, Hydroxide-{{Bonding Strength Measurements}}
  for {{Space}}-{{Based Optical Missions}}, International Journal of Applied
  Ceramic Technology 5~(4) (2008) 365--372.
\newblock \href {https://doi.org/10.1111/j.1744-7402.2008.02256.x}
  {\path{doi:10.1111/j.1744-7402.2008.02256.x}}.

\bibitem{BondingSiCtoSiC}
A.~Preston, G.~Mueller, Bonding sic to sic using a sodium silicate solution,
  International Journal of Applied Ceramic Technology 9~(4) (2012) 764--771.
\newblock \href
  {https://doi.org/https://doi.org/10.1111/j.1744-7402.2011.02644.x}
  {\path{doi:https://doi.org/10.1111/j.1744-7402.2011.02644.x}}.

\bibitem{matinfarReviewSodiumSilicate2023}
M.~Matinfar, J.~A. Nychka, A review of sodium silicate solutions:
  {{Structure}}, gelation, and syneresis, Advances in Colloid and Interface
  Science 322 (2023) 103036.
\newblock \href {https://doi.org/10.1016/j.cis.2023.103036}
  {\path{doi:10.1016/j.cis.2023.103036}}.

\bibitem{PhysRevB.84.195204}
A.~Dr\'eau, M.~Lesik, L.~Rondin, P.~Spinicelli, O.~Arcizet, J.-F. Roch,
  V.~Jacques, Avoiding power broadening in optically detected magnetic
  resonance of single nv defects for enhanced dc magnetic field sensitivity,
  Phys. Rev. B 84 (2011) 195204.
\newblock \href {https://doi.org/10.1103/PhysRevB.84.195204}
  {\path{doi:10.1103/PhysRevB.84.195204}}.

\bibitem{faraon_resonant_2011}
A.~Faraon, P.~E. Barclay, C.~Santori, K.-M.~C. Fu, R.~G. Beausoleil, Resonant
  enhancement of the zero-phonon emission from a colour centre in a diamond
  cavity, Nature Photonics 5~(5) (2011) 301--305.
\newblock \href {https://doi.org/10.1038/nphoton.2011.52}
  {\path{doi:10.1038/nphoton.2011.52}}.

\bibitem{Giakoumaki}
A.~Giakoumaki, G.~Coccia, V.~Bharadwaj~Shivakumar, J.~Hadden, A.~Bennett,
  B.~Sotillo, R.~Yoshizaki, P.~Olivero, O.~Jedrkiewicz, R.~Ramponi,
  S.~Pietralunga, M.~Bollani, A.~Bifone, P.~Barclay, A.~Kubanek, S.~Eaton,
  Quantum technologies in diamond enabled by laser processing, Applied Physics
  Letters 120 (2022) 020502.
\newblock \href {https://doi.org/10.1063/5.0080348}
  {\path{doi:10.1063/5.0080348}}.

\bibitem{mitchell_realizing_2019}
M.~Mitchell, D.~P. Lake, P.~E. Barclay, Realizing {Q} {\textgreater} 300 000 in
  diamond microdisks for optomechanics via etch optimization, APL Photonics
  4~(1) (2019) 016101.
\newblock \href {https://doi.org/10.1063/1.5053122}
  {\path{doi:10.1063/1.5053122}}.

\bibitem{ding_high-q_2024}
S.~W. Ding, M.~Haas, X.~Guo, K.~Kuruma, C.~Jin, Z.~Li, D.~D. Awschalom,
  N.~Delegan, F.~J. Heremans, A.~A. High, M.~Loncar, High-{Q} cavity interface
  for color centers in thin film diamond, Nature Communications 15~(1) (2024)
  6358.
\newblock \href {https://doi.org/10.1038/s41467-024-50667-5}
  {\path{doi:10.1038/s41467-024-50667-5}}.

\bibitem{guo_direct-bonded_2024}
X.~Guo, M.~Xie, A.~Addhya, A.~Linder, U.~Zvi, S.~Wang, X.~Yu, T.~D. Deshmukh,
  Y.~Liu, I.~N. Hammock, Z.~Li, C.~T. DeVault, A.~Butcher, A.~P. Esser-Kahn,
  D.~D. Awschalom, N.~Delegan, P.~C. Maurer, F.~J. Heremans, A.~A. High,
  Direct-bonded diamond membranes for heterogeneous quantum and electronic
  technologies, Nature Communications 15~(1) (2024) 8788.
\newblock \href {https://doi.org/10.1038/s41467-024-53150-3}
  {\path{doi:10.1038/s41467-024-53150-3}}.

\bibitem{katsumi_hybrid_2023}
R.~Katsumi, K.~Takada, S.~Naruse, K.~Kawai, D.~Sato, T.~Hizawa, T.~Yatsui,
  Hybrid integration of ensemble nitrogen-vacancy centers in single-crystal
  diamond based on pick-flip-and-place transfer printing, Applied Physics
  Letters 123~(11) (2023) 111108.
\newblock \href {https://doi.org/10.1063/5.0161268}
  {\path{doi:10.1063/5.0161268}}.

\end{thebibliography}

\end{document}